\documentclass[aps,showpacs,twocolumn]{revtex4}

\usepackage{amssymb}
\usepackage{amsmath,amssymb}
\usepackage{graphicx}

\begin{document}

\title{Fractional Vector Calculus and Fractional Special
Function}

\author{Ming-Fan Li}

\email{limf07@lzu.cn}

\author{Ji-Rong Ren}

\email{renjr@lzu.edu.cn}

\author{Tao Zhu}

\email{zhut05@lzu.cn}

\affiliation{Institute of Theoretical Physics, Lanzhou University,
Lanzhou, 730000, China}

\begin{abstract}

Fractional vector calculus is discussed in the spherical coordinate
framework. A variation of the Legendre equation and fractional
Bessel equation are solved by series expansion and numerically.
Finally, we generalize the hypergeometric functions.

\end{abstract}


\maketitle

\section{Introduction}

Fractional calculus is the calculus of differentiation and
integration of non-integer orders \cite{podlubny, kilbs}. During
last three decades, fractional calculus has gained much attention
due to its demonstrated applications in various fields of science
and engineering, such as anomalous diffusion \cite{scale},
fractional dynamical systems \cite{B-vanderPol, Brussel, happiness},
fractional quantum mechanics \cite{shroedinger}, fractional
statistics and thermodynamics \cite{entropy}, to name a few.

Fractional vector calculus (FVC) is important in describing
processes in fractal media, fractional electrodynamics and
fractional hydrodynamics \cite{poisson,fvc}. But an effective FVC is
still lacking. There are many problems in defining an effective FVC.
One is that fractional integral and fractional derivative are
defined ``half" , that is to say, they are defined only on the right
or the left side of an initial point. And if we make fractional
series expansion, functions are all expanded on the right or the
left neighborhood of the initial point. We cannot across this
cutting point. So if we want to describe the behavior near the
initial point, we need define both the right and the left functions.
The situation will become even more complicated when we deal with
high dimensions. In this letter we will define FVC in spherical
coordinate framework. Since in the spherical coordinate framework
the radius is naturally bounded to the positive half, we need just
one fractional derivative.

Using this frame, we will discuss the Laplacian equation \cite{yzj}
with fractional radius derivative in 3-d space and the heat conduct
equation \cite{yzj} in 2-d space with cylindrical symmetry. As a
result, the corresponding special functions will be generalized.
Finally, we will generalize hypergeometric functions
\cite{wzx,wiki_HyperF,wiki_ConfHyperF}.

\section{Fractional calculus}\label{FC}

To start, let's briefly recall some basic facts in fractional
calculus \cite{podlubny, kilbs}. There are many ways to define
fractional integral and fractional derivative. In this letter we
will use Riemann-Liouville fractional integral and Caputo fractional
derivative.

Let $f(x)$ be a function defined on the right side of $a$. Let
$\alpha$ be a positive real. The Riemann-Liouville fractional
integral is defined by
\begin{equation}\label{}
    I^{\alpha}f(x)=\frac{1}{\Gamma(\alpha)}\int_a^x
    (x-\xi)^{\alpha-1}f(\xi)~d\xi.
\end{equation}
The integral has a memory kernel.

Let $A\equiv [\alpha]+1$. The Caputo fractional derivative is
defined by
\begin{eqnarray}\label{}
    D^{\alpha}f(x) &=& I^{A-\alpha}\frac{d^A}{dx^A}f(x)\nonumber\\
                   &=& \frac{1}{\Gamma(A-\alpha)}\int_a^x(x-\xi)^{A-\alpha-1}\frac{d^A}{d\xi^A}f(\xi)~d\xi.
\end{eqnarray}

They have the following properties:
\begin{eqnarray}\label{relation1}
    D^{\alpha}(x-a)^{\beta} = 0,~~~~~~~~~~~~~~~~~~~~~~~~\beta\in \{0,1,...,[\alpha]\}; \\
    D^{\alpha}(x-a)^{\beta} = \frac{\Gamma(\beta+1)}{\Gamma(\beta-\alpha+1)}(x-a)^{\beta-\alpha}, \beta~\text{otherwise};  \\
    I^{\alpha}(x-a)^{\beta} =
    \frac{\Gamma(\beta+1)}{\Gamma(\beta+\alpha+1)}(x-a)^{\beta+\alpha}.~~~~~~~~~~~~~~~~~
\end{eqnarray}
These properties are just fractional generalizations of
\begin{equation}\label{dpower}
    \frac{d^n}{d x^n}x^m = \frac{m!}{(m-n)!}x^{m-n},~~~~n\in N, ~~m\neq
    0,
\end{equation}
\begin{equation}
    \int_0^x dx~x^m = \frac{m!}{(m+1)!}x^{m+1}.
\end{equation}

For a `good' function, one can define its fractional Taylor series
\begin{equation}\label{}
    f(x)=\sum_{m=0}^{\infty} (D^{\alpha})^m f(x)\big|_{x=a}\cdot [(I^{\alpha})^m\cdot 1].
\end{equation}
Explicitly,
\begin{equation}\label{}
    (I^{\alpha})^m\cdot 1=\frac{1}{\Gamma(m\alpha+1)}(x-a)^{m\alpha}.
\end{equation}

\section{Fractional vector calculus}\label{FVC}

As has been aforementioned, fractional derivative is defined only on
the right or the left half of the real line, which gives
complications in defining an effective FVC with Cartesian
coordinates. So it may be more feasible to do FVC with spherical
coordinates.

\subsection{Spherical coordinates}

The spherical coordinates of 3-dimension space is a triplet $(r,
\theta, \phi)$. In one-order calculus, the gradient of a scale
function $u(r, \theta, \phi)$ is
\begin{equation}\label{}
    \textbf{Grad}~u = \bigg(\textbf{e}_r \frac{\partial}{\partial r}+
    \textbf{e}_{\theta}\frac{1}{r}\frac{\partial}{\partial\theta}+
    \textbf{e}_{\phi}\frac{1}{r\text{sin}\theta}\frac{\partial}{\partial\phi}\bigg)u.
\end{equation}

We generalize this definition to fractional calculus as
\begin{equation}\label{}
    \textbf{Grad}^{\alpha}~u = \bigg(\textbf{e}_r D^{\alpha}_r+
    \textbf{e}_{\theta}\frac{\Gamma(\alpha+1)}{r^{\alpha}}\frac{\partial}{\partial\theta}+
    \textbf{e}_{\phi}\frac{\Gamma(\alpha+1)}{r^{\alpha}\text{sin}^{\alpha}\theta}\frac{\partial}{\partial\phi}\bigg)u,
\end{equation}
where $D^{\alpha}_r ~u=\frac{1}{\Gamma(A-\alpha)}\int_0^r
    (r-\rho)^{A-\alpha-1}\frac{d^A}{d\rho^A}u(\rho, \theta, \phi)d\rho$.
For anisotropic space, $\alpha$ is a function of $\theta$ and
$\phi$. For isotropic space, $\alpha$ is a constant and independent
of $\theta$ and $\phi$. We will just consider the isotropic case.

\begin{figure}
  \includegraphics[width=7.8cm]{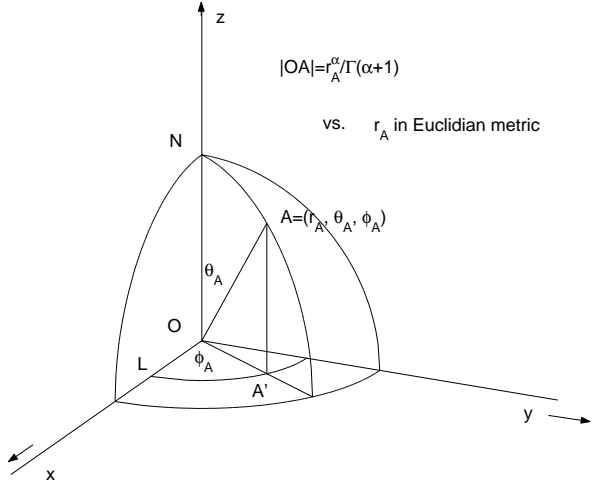}\\
  \caption{Spherical framework with fractional radius derivative. $|OA|=\frac{1}{\Gamma(\alpha+1)}r_A^{\alpha}$, $\widehat{NA}=\frac{1}{\Gamma(\alpha+1)}r_A^{\alpha}\theta_A$, and $\widehat{LA'}=\frac{1}{\Gamma(\alpha+1)}r_A^{\alpha}\text{sin}^{\alpha}\theta_A~\phi_A$.}\label{spherical}
\end{figure}

By this definition, the real space metric is changed to an effective
metric. This can be easily seen from Fig. \ref{spherical}. The
radius $|OA|$ now is $[I^{\alpha}_r\cdot
1]|_{r=0}^{r=r_A}=\frac{1}{\Gamma(\alpha+1)}r_A^{\alpha}$; the arc
length $\widehat{NA}=\frac{1}{\Gamma(\alpha+1)}r_A^{\alpha}\theta_A$
and the arc length
$\widehat{LA'}=\frac{1}{\Gamma(\alpha+1)}r_A^{\alpha}\text{sin}^{\alpha}\theta_A~\phi_A$.
This kind of metric is not addictive since $|OB|\neq|OA|+|AB|$ even
when $O$, $A$ and $B$ are on the same straight line. This is due to
the non-locality of the fractional operations.

\begin{widetext}
By the above generalization, the divergence of a vector function
$\textbf{A}=(A_r, A_{\theta}, A_{\phi})$ is
\begin{equation}\label{}
    \text{div}^{\alpha}~\textbf{A} =
    \frac{1}{r^{2\alpha}\text{sin}^{\alpha}\theta}D^{\alpha}_r(r^{2\alpha}\text{sin}^{\alpha}\theta
    A_r)+\frac{\Gamma(\alpha+1)}{r^{2\alpha}\text{sin}^{\alpha}\theta}\frac{\partial}{\partial\theta}(r^{\alpha}\text{sin}^{\alpha}\theta
    A_{\theta})+\frac{\Gamma(\alpha+1)}{r^{2\alpha}\text{sin}^{\alpha}\theta}\frac{\partial}{\partial\phi}(r^{\alpha}A_{\phi}).
\end{equation}

The Laplacian of a scale function $u(r, \theta, \phi)$ will be
\begin{eqnarray}
  \triangle^{\alpha} ~u &\equiv& \text{div}^{\alpha}\textbf{Grad}^{\alpha} ~u \nonumber\\
                             &=& \frac{1}{r^{2\alpha}}D^{\alpha}_r(r^{2\alpha}D^{\alpha}_r ~u) +\frac{\Gamma^2(\alpha+1)}{r^{2\alpha}\text{sin}^{\alpha}\theta}\frac{\partial}{\partial\theta}(\text{sin}^{\alpha}\theta\frac{\partial}{\partial\theta} ~u) +\frac{\Gamma^2(\alpha+1)}{r^{2\alpha}\text{sin}^{2\alpha}\theta}\frac{\partial^2}{\partial\phi^2}
                             ~u.
\end{eqnarray}
\end{widetext}

The rotor operator can be defined as well. Since in this letter we
do not deal with the rotor, its definition will not be given.

\subsection{Polor coordinates}

Likewise, the fractional gradient operator of two dimensions in
polor coordinates can be defined as
\begin{equation}\label{}
    \textbf{Grad}^{\alpha}~u = \bigg(\textbf{e}_r D^{\alpha}_r+
    \textbf{e}_{\theta}\frac{\Gamma(\alpha+1)}{r^{\alpha}}\frac{\partial}{\partial\theta}\bigg) u.
\end{equation}

The divergence is
\begin{equation}\label{}
    \text{div}^{\alpha}~\textbf{A} =
    \frac{1}{r^{\alpha}}D^{\alpha}_r(r^{\alpha}A_r)+\frac{\Gamma(\alpha+1)}{r^{\alpha}}\frac{\partial}{\partial\theta}A_{\theta}.
\end{equation}

The Laplacian of a scale function $u(r, \theta)$ is
\begin{eqnarray}
  \triangle^{\alpha} ~u &\equiv& \text{div}^{\alpha}\textbf{Grad}^{\alpha} ~u \nonumber\\
                             &=& \frac{1}{r^{\alpha}}D^{\alpha}_r(r^{\alpha}D^{\alpha}_r ~u)+\frac{\Gamma^2(\alpha+1)}{r^{2\alpha}}\frac{\partial^2}{\partial\theta^2}u.
\end{eqnarray}

\section{Fractional spherical equation and fractional cylindrical equation}

In this section, we consider the Laplacian equation with the 3-d
Laplacian operaror defined above and the fractional heat conduct
equation in a 2-d space with cylindrical symmetry.

\subsection{Fractional Laplacian equation}
With the Laplacian operator defined above, the Laplacian equation
becomes
\begin{widetext}
\begin{equation}\label{}
    \triangle^{\alpha} ~u =\frac{1}{r^{2\alpha}}D^{\alpha}_r(r^{2\alpha}D^{\alpha}_r ~u) +\frac{\Gamma^2(\alpha+1)}{r^{2\alpha}\text{sin}^{\alpha}\theta}\frac{\partial}{\partial\theta}(\text{sin}^{\alpha}\theta\frac{\partial}{\partial\theta} ~u) +\frac{\Gamma^2(\alpha+1)}{r^{2\alpha}\text{sin}^{2\alpha}\theta}\frac{\partial^2}{\partial\phi^2}
                             ~u=0.
\end{equation}
\end{widetext}

This equation can be solved by separation of variables. Let
$u=R(r)\Theta(\theta)\Phi(\phi)$ and substitute, the result is
\begin{equation}\label{}
   \frac{1}{\Theta\text{sin}^{\alpha}\theta}\frac{d}{d\theta}\bigg(\text{sin}^{\alpha}\theta\frac{d\Theta}{d\theta}\bigg)+\frac{1}{\Phi\text{sin}^{2\alpha}\theta}\frac{d^2\Phi}{d\phi^2}=-\lambda,
\end{equation}
\begin{equation}\label{}
    \frac{1}{R}D^{\alpha}_r(r^{2\alpha}D^{\alpha}_r R)=\lambda\Gamma^2(\alpha+1).
\end{equation}

The second equation can be solved by fractional series expansion.
Let $R=\sum_{m=-\infty}^{\infty}c_m r^{m\alpha}$, and substitute,
the result is that the nonzero components are the ones with $m$
satisfying
\begin{equation}\label{alpha_m_l}
    \frac{\Gamma(m\alpha+\alpha+1)}{\Gamma(m\alpha-\alpha+1)}=\lambda\Gamma^2(\alpha+1).
\end{equation}
Otherwise, $c_m$=0.

The first equation is a variation of the ordinary spherical harmonic
equation. By further decomposition, it turns to
\begin{equation}\label{}
    \frac{d^2\Phi}{d\phi^2}+m^2\Phi=0, ~~~~~(m=0, 1, 2, 3, ...)
\end{equation}
\begin{equation}\label{}
   \frac{\text{sin}^{\alpha}\theta}{\Theta}\frac{d}{d\theta}\bigg(\text{sin}^{\alpha}\theta\frac{d\Theta}{d\theta}\bigg)=m^2-\lambda\text{sin}^{2\alpha}\theta.
\end{equation}

The first equation is simple. The second equation can be transformed
by changing variables $x=\text{cos}\theta$ and $p(x)=\Theta(\theta)$
to
\begin{equation}\label{}
    (1-x^2)\frac{d^2p}{dx^2}-(1+\alpha)x\frac{dp}{dx}+\bigg[\lambda-\frac{m^2}{(1-x^2)^{\alpha}}\bigg]p=0.
\end{equation}
This is a variation of the ordinary associated Legendre equation
\cite{yzj,wzx}. By setting $\alpha=1$, we recover the ordinary
associated Legendre equation.

\begin{figure}
  \includegraphics[width=7.8cm]{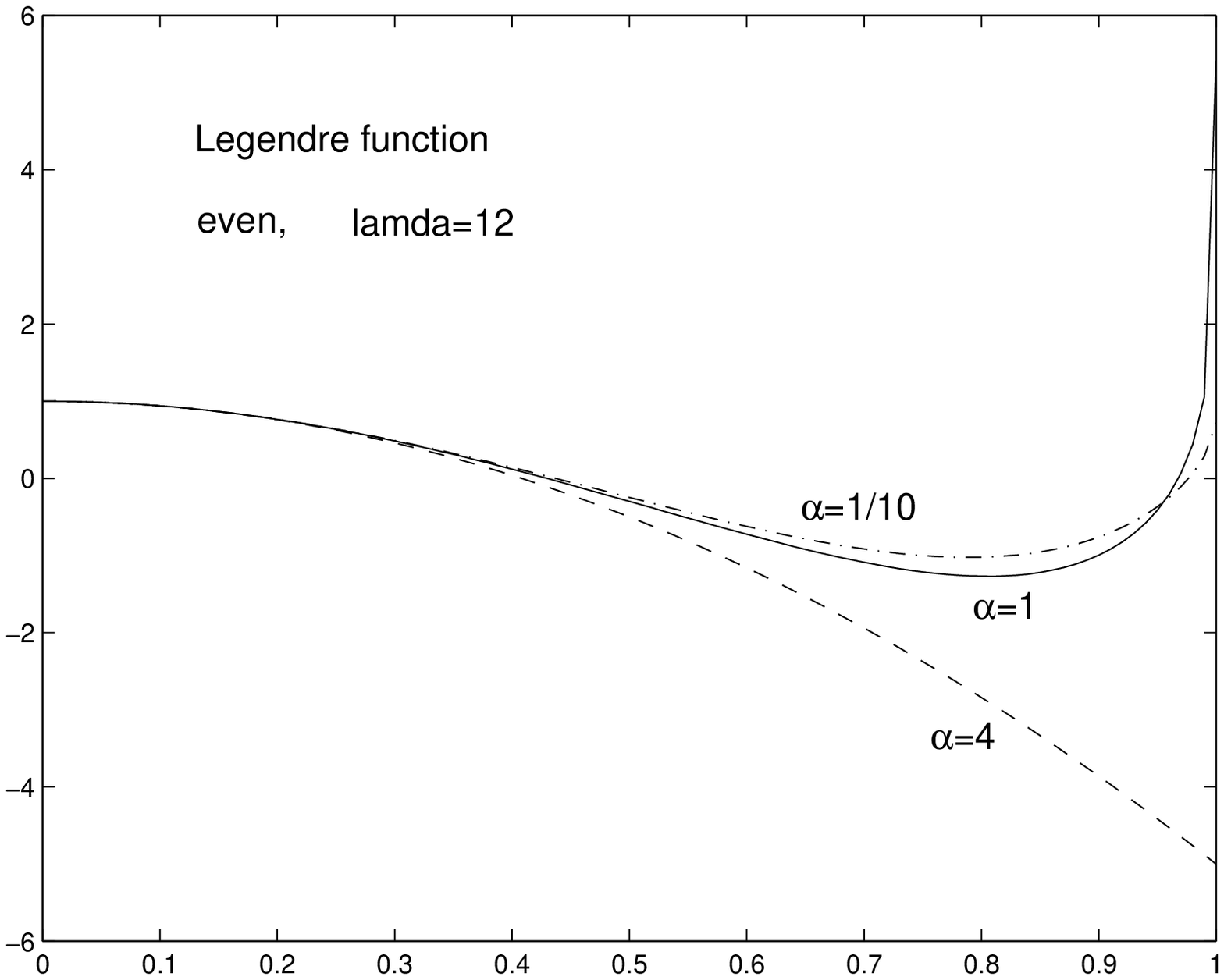}\\
  \caption{Legendre even function. Since the fractional index $\alpha$ occurs as a multiplicative factor in the variation version of the Legendre equation, small difference from 1 will not make large changes to the profile. We calculated the function with other small difference $\alpha$'s, but the results are not shown for their curves are very close to each other. Notice the direct of the curves.}\label{legendre_even}
\end{figure}

\begin{figure}
  \includegraphics[width=7.8cm]{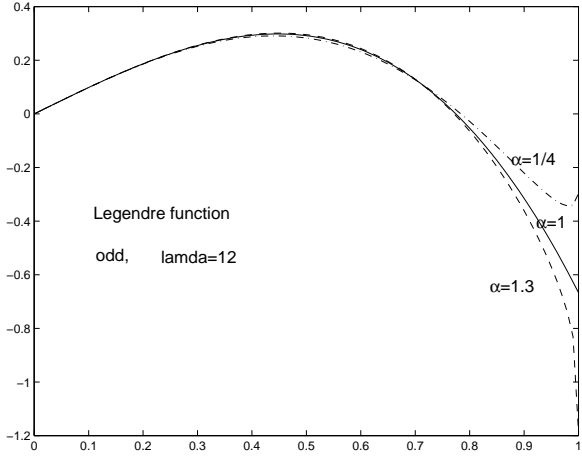}\\
  \caption{Legendre odd function. Notice the direct of the curves.}\label{legendre_odd}
\end{figure}

When $m=0$ it is the Legendre equation. Make the ansatz
$p(x)=\sum_{n=0}^{\infty}c_nx^n$ and substitute, we find that the
terms with even powers of $x$ and the terms with odd powers of $x$
are independent, so we can write $p(x)=c_0\cdot
p_{\text{even}}(x)+c_1\cdot p_{\text{odd}}(x)$. The relation of
successive coefficients is
\begin{equation}\label{}
    c_{n+2}=c_{n}\frac{n^2+n\alpha-\lambda}{(n+2)(n+1)}.
\end{equation}
Let $\alpha=1$ and $\lambda=l(l+1)$, we will get the Legendre
polynomials. We calculated numerically the functions with some
different values of the parameters and displayed the results in Fig.
\ref{legendre_even} and Fig. \ref{legendre_odd}.


\subsection{Fractional cylindrical equation}

The heat conduct equation in 2-d space is
\begin{equation}\label{}
    \frac{\partial u}{\partial t}=a^2\triangle_2 u.
\end{equation}
Here $\triangle_2$ is the two dimensional Laplacian operator. Using
the fractional cylindrical Laplacian operator defined above, this
equation becomes
\begin{equation}\label{}
    \frac{\partial u}{\partial t} = \frac{1}{r^{\alpha}}D^{\alpha}_r(r^{\alpha}D^{\alpha}_r ~u)+\frac{\Gamma^2(\alpha+1)}{r^{2\alpha}}\frac{\partial^2}{\partial\theta^2}u.
\end{equation}

By separation of variables $u(t, r, \theta)=R(r)\Theta(\theta)T(t)$,
it can be decomposed to
\begin{equation}\label{}
    T'+a^2k^2T=0,
\end{equation}
\begin{equation}\label{}
    \frac{\partial^2\Theta}{\partial\theta^2}+\nu^2\Theta=0,
\end{equation}
\begin{equation}\label{}
    \frac{1}{\Gamma^2(\alpha+1)}r^{\alpha}D^{\alpha}_r(r^{\alpha}D^{\alpha}_r
    ~R)+k^2\frac{r^{2\alpha}}{\Gamma^2(\alpha+1)}R-\nu^2R=0.
\end{equation}

The first two equations are simple. The third equation is a
fractional generalization of the Bessel equation \cite{yzj,wzx}. It
can be solved by fractional series expansion. Since Bessel equation
is singular at $r=0$. We must use such ansatz:
$R=r^{\alpha\rho}\sum_{m=0}^{\infty}c_mr^{\alpha m}$. Substitute it
into the above equation, we get
\begin{equation}\label{}
    c_0\bigg[\bigg(\frac{\Gamma(\alpha\rho+1)}{\Gamma(\alpha\rho-\alpha+1)}\bigg)^2-\nu^2\Gamma^2(\alpha+1)\bigg]=0,
\end{equation}
\begin{equation}\label{}
    c_1\bigg[\bigg(\frac{\Gamma(\alpha\rho+\alpha+1)}{\Gamma(\alpha\rho+1)}\bigg)^2-\nu^2\Gamma^2(\alpha+1)\bigg]=0,
\end{equation}
\begin{equation}\label{}
    c_m\bigg[\bigg(\frac{\Gamma(\alpha\rho+\alpha m+1)}{\Gamma(\alpha\rho+\alpha
    m-\alpha+1)}\bigg)^2-\nu^2\Gamma^2(\alpha+1)\bigg]+c_{m-2}k^2=0.
\end{equation}

To have a starting term, $c_0\neq 0$, so
\begin{equation}\label{nu_rho}
    \bigg[\bigg(\frac{\Gamma(\alpha\rho+1)}{\Gamma(\alpha\rho-\alpha+1)}\bigg)^2-\nu^2\Gamma^2(\alpha+1)\bigg]=0,
\end{equation}
and $c_1=0$.

By the recursive relation, a solution is implied
\begin{equation}\label{}
    R_{\rho}(r)=r^{\alpha\rho}\sum_{n=0}^{\infty}(-1)^nd_nk^{2n}r^{\alpha\cdot2n},
\end{equation}
where $d_0=1$,
\begin{equation}\label{}
    d_n=d_{n-1}\frac{1}{\bigg[\bigg(\frac{\Gamma(\alpha\rho+\alpha\cdot2n+1)}{\Gamma(\alpha\rho+\alpha\cdot2n-\alpha+1)}\bigg)^2-\nu^2\Gamma^2(\alpha+1)\bigg]},
\end{equation}
and $\rho$ satisfies Eq.(\ref{nu_rho}).

\begin{figure}
  \includegraphics[width=7.8cm]{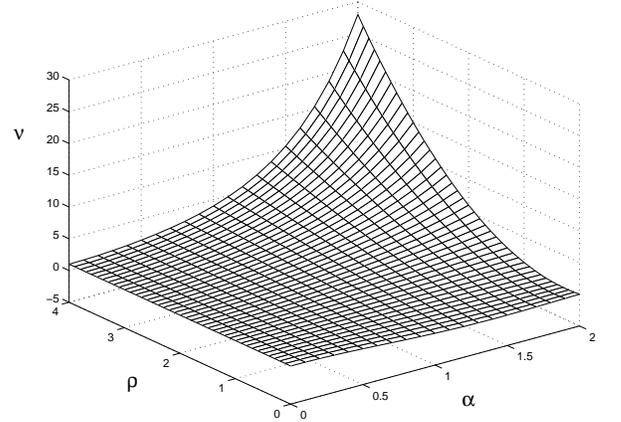}\\
  \caption{Parameters of fractional Bessel function. $\nu$ as a function of $\alpha$ and $\rho$.}\label{bessel_nu}
\end{figure}

\begin{figure}
  \includegraphics[width=7.8cm]{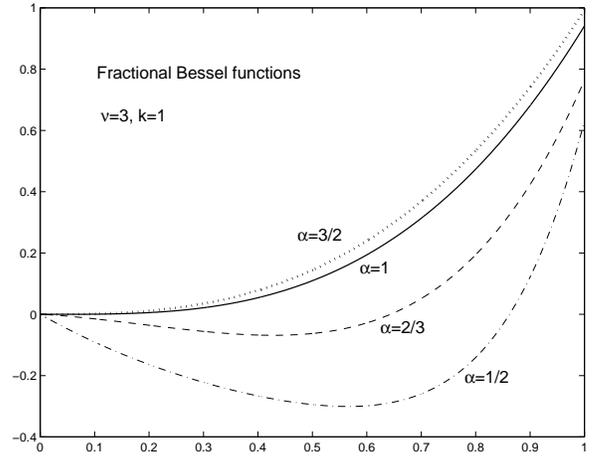}\\
  \caption{Fractional Bessel function. The fractional index $\alpha$ shows up in the exponentials of the fractional Bessel series; a small difference from 1 changes the profile largely. For a big $\alpha$, $\rho$ is small, so the change is suppressed.}\label{bessel}
\end{figure}

Eq.(\ref{nu_rho}) in one-order calculus ($\alpha=1$) is simply
$\rho=\pm \nu$. We meshed in Fig. \ref{bessel_nu} the surface
defined by the equation (\ref{nu_rho}). After solving the equation
with $\nu=3$ for $\rho$ when $\alpha$ varies, we plotted in Fig.
\ref{bessel} $R_{\rho}(r)$ belonging to different values of
$\alpha$.


\section{Fractional hypergeometric function}\label{FHyperF}

There are other types of special functions in mathematical physics.
A most famous one is the hypergeometric function
\cite{wzx,wiki_HyperF,wiki_ConfHyperF}. In this section, we will try
to define a fractional generalization of the hypergeometric
functions.

Let's first consider the generalization of the confluent
hypergeometric differential equation:
\begin{equation}\label{}
    z^{\alpha}(D^{\alpha})^2
    y+(c-z^{\alpha})D^{\alpha}y-ay=0.
\end{equation}
Here $a$ and $c$ are complex parameters. When $\alpha=1$, this is
the ordinary confluent hypergeometric equation.

\begin{figure}
  \includegraphics[width=7.8cm]{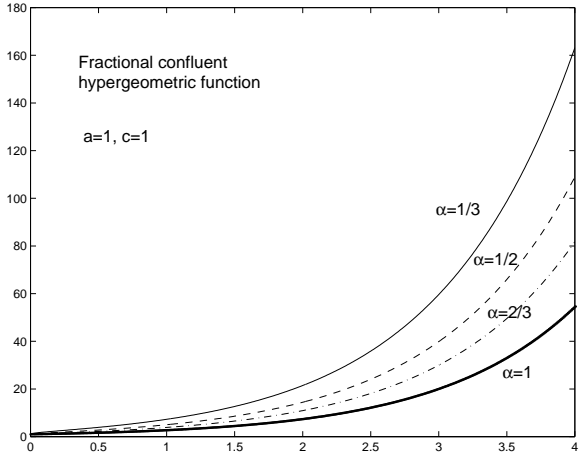}\\
  \caption{Fractional confluent hypergeometric function.}\label{conf_hyper}
\end{figure}

Introducing the fractional Taylor series
\begin{equation}\label{yFTS}
    y(z)=\sum_{k=0}^{\infty}c_k z^{\alpha\cdot k},
\end{equation}
and substituting, we get the ratio of successive coefficients
\begin{eqnarray}
  \frac{c_{k+1}\cdot\Gamma(k\alpha+\alpha+1)}{c_{k}\cdot\Gamma(k\alpha+1)} &=& \frac{a+\frac{\Gamma(k\alpha+1)}{\Gamma(k\alpha-\alpha+1)}}{c+\frac{\Gamma(k\alpha+1)}{\Gamma(k\alpha-\alpha+1)}},\\
  \frac{c_1\cdot\Gamma(\alpha+1)}{c_0} &=& \frac{a}{c}.
\end{eqnarray}
Thus we get a solution of the above differential equation,
\begin{equation}\label{fconfhyperf}
    y(z)=\sum_{k=0}^{\infty}\frac{(a)^{\alpha}_k}{(c)^{\alpha}_k}\frac{1}{\Gamma(k\alpha+1)}z^{\alpha\cdot k}.
\end{equation}
Here $(a)^{\alpha}_k$ is defined as
\begin{eqnarray}
  (a)^{\alpha}_0 &=& 1, ~~~~~(a)^{\alpha}_1=a, \nonumber\\
  (a)^{\alpha}_k &=& (a)^{\alpha}_1\bigg(a+\frac{\Gamma(\alpha+1)}{\Gamma(1)}\bigg)...\bigg(a+\frac{\Gamma(k\alpha-\alpha+1)}{\Gamma(k\alpha-2\alpha+1)}\bigg),\nonumber\\
  && ~~~~~~~~~~~~~~~~~~~~~~~~~~~~~~~~~~~~~~~~~~~~~~~k\geq2.
\end{eqnarray}
This can be seen as a fractional generalization of the rising
factorial
\begin{equation}\label{risingfactorial}
    (a)_k=a(a+1)...(a+k-1).
\end{equation}
And the series (\ref{fconfhyperf}) can be seen as a fractional
generalization the confluent hypergeometric function. If $\alpha=1$,
it is exactly the confluent hypergeometric function. Profiles of
this series (\ref{fconfhyperf}) with different values of $\alpha$
are displayed in Fig. \ref{conf_hyper}.

For the fractional Gauss hypergeometric function, consider the
following series
\begin{equation}\label{fhyperf}
    y(z)=\sum_{k=0}^{\infty}\frac{(a)^{\alpha}_k(b)^{\alpha}_k}{(c)^{\alpha}_k}\frac{1}{\Gamma(k\alpha+1)}z^{\alpha\cdot
    k},
\end{equation}
which reduces to the Gauss hypergeometric series when $\alpha=1$.

The ratio of successive coefficients is
\begin{equation}\label{}
    \frac{c_{k+1}\cdot\Gamma(k\alpha+\alpha+1)}{c_{k}\cdot\Gamma(k\alpha+1)} =
    \frac{\bigg(a+\frac{\Gamma(k\alpha+1)}{\Gamma(k\alpha-\alpha+1)}\bigg)\bigg(b+\frac{\Gamma(k\alpha+1)}{\Gamma(k\alpha-\alpha+1)}\bigg)}{\bigg(c+\frac{\Gamma(k\alpha+1)}{\Gamma(k\alpha-\alpha+1)}\bigg)},
\end{equation}
or
\begin{widetext}
\begin{eqnarray}\label{fGHf_Coeff}
  &&c_{k+1}\cdot
    c\frac{\Gamma(k\alpha+1)}{\Gamma(k\alpha-\alpha+1)}+c_{k+1}\cdot
    \frac{\Gamma(k\alpha+\alpha+1)}{\Gamma(k\alpha-\alpha+1)} \nonumber \\
  &&=c_k\cdot
    ab+c_k\cdot(a+b)\frac{\Gamma(k\alpha+1)}{\Gamma(k\alpha-\alpha+1)}+c_k\cdot\frac{\Gamma(k\alpha+1)}{\Gamma(k\alpha-\alpha+1)}\frac{\Gamma(k\alpha+1)}{\Gamma(k\alpha-\alpha+1)}.
\end{eqnarray}
\end{widetext}

\begin{figure}
  \includegraphics[width=7.8cm]{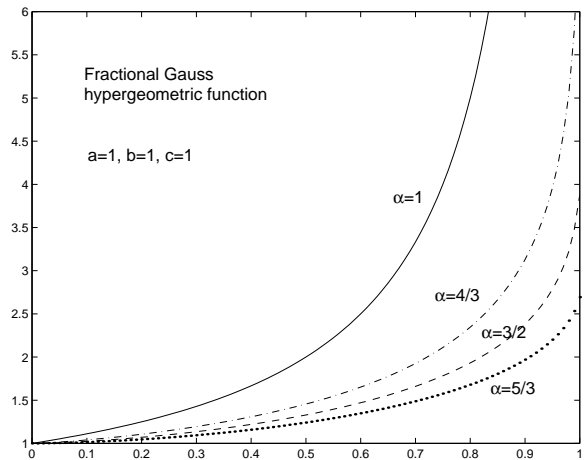}\\
  \caption{Fractional Gauss hypergeometric function. Gauss hypergeometric function is divergent at 1. From the figure we can see this is not the case for fractional equivalences with a bigger $\alpha$.}\label{Gauss_hyper}
\end{figure}

Since
\begin{equation}
  y(z) = \sum_{k=0}^{\infty}c_k z^{\alpha\cdot k},
\end{equation}
\begin{equation}
  z^{\alpha}D^{\alpha}y(z) = \sum_{k=1}^{\infty}c_k \frac{\Gamma(k\alpha+1)}{\Gamma(k\alpha-\alpha+1)} z^{\alpha\cdot
  k},
\end{equation}
the equation (\ref{fGHf_Coeff}) can be translated to a fractional
differential equation
\begin{eqnarray}
  && ab\cdot y(z)+(a+b)z^{\alpha}D^{\alpha}y(z)+z^{\alpha}D^{\alpha}\big[z^{\alpha}D^{\alpha}y(z)\big] \nonumber \\
  && ~~~~~~~~~~~~~~~~~~=c\cdot D^{\alpha}y(z)+z^{\alpha}(D^{\alpha})^2
  y(z).
\end{eqnarray}
When $\alpha=1$, this equation reduces to the ordinary Gauss
hypergeometric equation. We draw curves of some example functions in
Fig. \ref{Gauss_hyper}.

\section{Summary}\label{Summary}

In this letter, we defined fractional vector calculus in the
spherical coordinate framework. We discussed Laplacian equation and
heat conduct equation in this kind of framework. Some special
functions are generalized.

The geometry induced by fractional operations is non-addictive and
nonlocal. This kind of geometry is not Riemannian; since fractional
calculus has found applications in many areas of science and
engineering, it is the effective geometry of such physics processes.
It will be meaningful to investigate further this kind of geometry.

For a complete fractional vector calculus, fractional vector
integral is indispensable. But in this letter we restrained
ourselves from this direction. It will be interesting to discuss
fractional vector integral in our framework and generalize Green's,
Stokes' and Gauss's theorem.

Special functions show up in different areas of physics and
engineering. A fractional generalization of these functions may find
applications in similar situations with anomalous tailing behaviors
and/or nonlocal properties.

\end{document}